# Chemical Responsive Single Crystal Organic Magnet

Yan Cui,[1,2] Huiping Zhu,[1,2] Lei Wang[1,2], Bo Li ,[1,2] Zhengsheng Han,[1,2] and Jiajun Luo[1,2,a)]

[1]Institute of Microelectronics, Chinese Academy of Sciences, Beijing 100029, China
[2]Key Laboratory of Silicon Device Technology, Chinese Academy of Sciences, Beijing 100029, China

**ABSTRACT:**

The flexibility and diversity of organic chemistry have yielded many materials in which magnetism can be varied. However, most methods used for changing magnetism are inefficient or destructive to the magnetic material. Here we report high-performance magnetic control of a gas-responsive single-molecule magnet (SMM). The results exhibit that the magnetic properties of the SMM can be significantly changed according to the gas environment it is in and some of the magnetic states can be reversibly transformed or coexistent in the SMM through artificial control. More importantly, the single crystalline structure of the SMM remains unchanged during the transformation process except for slight change of the lattice constant. Thus, this work opens up new insights into the stimuli-responsive magnetic materials which have great prospects for application in artificial design magnetic network and also highlight their potential as smart materials.

Materials that change their magnetic properties in response to the external stimuli have long been of interest for their potential applicability in magnetic storage device, spintronics and smart magnetic materials[1-5]. Organic materials are suitable candidates for such materials due to their chemical diversity, flexibility and designability. However, the most common methods[6-17] to tune the magnetism of organic materials provide poor controllability and even destroy the materials. For these reasons, finding a proper tunable magnetic material and method has become a challenge for the scientific community. One promising candidate is a chemical responsive organic magnet in which the component associated with the magnetism can be flexibly substituted with very little damage to the crystal structure in response to external chemical stimuli.

Here we report a gas-responsive single-crystal single-molecule magnet (SMM) [$Mn_3O(Et_3\text{-}sao)_3(ClO_4)(MeOH)_3$][18,19,20] (hereafter $Mn_3$-$CH_3OH$ for short, because it has methanol as a ligands). $Mn_3$-$CH_3OH$ complex crystallizes in the trigonal system, space group of R-3 reported by Inglis *et al*. It has a honeycomb-like magnetic network and exhibits slow relaxation and quantum tunneling of magnetization (QTM) at low temperature due to its high anisotropy energy barrier and large ground state spin[21,22]. Distinct from isolated SMMs system[23] and dimer SMMs system[24], $Mn_3$-$CH_3OH$ has identical intermolecular antiferromagnetic (AFM) exchange interaction in its honeycomb-like magnetic network[18] and the AFM exchange interaction depends on the intermolecular hydrogen bond formed by the ligand on the $Mn_3$-$CH_3OH$ molecule[18,19,20]. Recently, some of us has reported that the methanol ligands can be replaced when the $Mn_3$-$CH_3OH$ crystal is exposed in the air for a few days[25,26,27], while the space group and the crystal symmetry remain unchanged[18,25,26]. The change of methanol ligands induces a significant change of the magnetism, such as AFM exchange interaction and the QTM effect. Interestingly, the methanol ligands can be reversibly grafted back to the molecule when the sample is exposed in methanol gas[25]. However, the nature of the reaction has not been verified in the previous work.

In order to understand the chemical mechanism of the reaction and employ it to achieve control of magnetism without destroying the crystal structure, we carried out more detailed study on the transformation of $Mn_3$-$CH_3OH$. The prototype of $Mn_3$-$CH_3OH$ complex is synthesized according to the method reported by Inglis *et al*[18]. The single crystal picture is shown in Fig.1a. The intermolecular hydrogen bond[18] which associates with the AFM exchange interaction is formed by the H atom on the methanol ligand and the O atom on the neighboring molecules, leading to a two-dimensional honeycomb-like magnetic network (in ab plane) as shown in Fig.1b and Fig.1c. In the c direction (easy axis), the intermolecular interaction is van der Waals' force, thus $Mn_3$-$CH_3OH$ can be regarded as a two-dimensional magnetic system.

First, the synthesized samples of $Mn_3$-$CH_3OH$ from the same batch are divided into four groups and store in air, oxygen dried by sulphuric acid, water vapor and water vapor with a certain concentration of oxygen for about two weeks, respectively. We find that a ligand displacement reaction take place. The methanol ligands on $Mn_3$-$CH_3OH$ are replaced by water molecule for samples store in air, water vapor and water vapor with oxygen, while samples in the oxygen dried by sulphuric acid keep unchanged (the sample with water ligands is named as $Mn_3$-$H_2O$, the crystal structure of it is available free of charge via www.ccdc.cam.ac.uk/data_request/cif with the number CCDC-1475184). And samples store in environment with oxygen change faster. That indicates the oxygen plays a role of catalyst during the reaction.

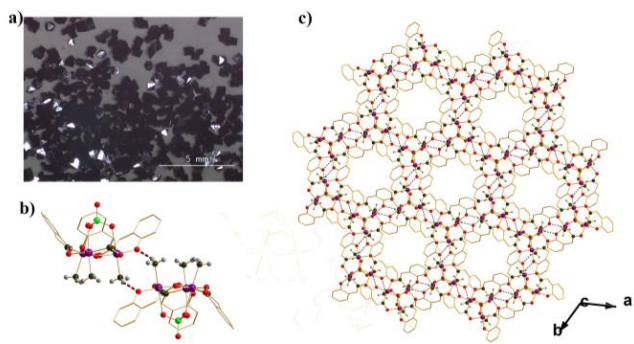

**Figure 1.** (a) The picture of the single-crystal sample of the three kinds of $Mn_3$. (b) The hydrogen bond between $Mn_3$-$NH_3$ molecules. (c) The ab plane structure of $Mn_3$-$NH_3$. The three kinds of $Mn_3$ have the same lattice structure. Colour code: Mn, purple; O, red; C, black; N, olive green; Cl, green; H, cyan. Most carbon and hydrogen atoms are neglected for clarity.

Based on this conclusion, we first propose a gas-responsive scheme to attain the goal of non-destructive magnetism control of SMM $Mn_3$. Both methanol ligands and water ligands link with Mn ions via Mn-O coordination bond, which is relatively weaker than other kinds of coordination bonds such as Mn-N coordination bond. Hence, the methanol ligands and the water ligands should be changed to some kind of ligands contain N when $Mn_3$-$CH_3OH$ and $Mn_3$-$H_2O$ are placed in a suitable nitrogen-rich atmosphere. In consideration of the fact that the size of the gas molecule should be small enough to infiltrate inside the crystal, we choose ammonia for the displacement reaction. We exposed the single-crystal of $Mn_3$-$CH_3OH$ and $Mn_3$-$H_2O$ in the atmosphere of dried ammonia gas. The dried ammonia gas is obtained by warming up 30 mL aqueous ammonia at 40 °C and then dried by soda lime. The results indicate that the methanol ligands on $Mn_3$-$CH_3OH$ and water ligands on $Mn_3$-$H_2O$ are indeed replaced by ammonia ligands (we name the new molecule as $Mn_3$-$NH_3$, the molecular structure is shown in Fig.2). The whole transmission process is shown in Fig.2. What's more, the ligands displacement rate significantly increases in the presence of oxygen. As expected, the single-crystal structure of $Mn_3$-$NH_3$ was not destroyed during the reaction. The single-crystal X-ray diffraction (SXRD) results exhibit that the $Mn_3$-$NH_3$ still crystallizes in the trigonal system, space group R-3 while the lattice constants show a slight variation[18,25] (the crystal structure of it is available via www.ccdc.cam.ac.uk/data_request/cif with the number CCDC-1475183). On the other hand, the $Mn_3$-$NH_3$ cannot be changed back to $Mn_3$-$CH_3OH$ and $Mn_3$-$H_2O$ when exposed to methanol gas and vapor, indicating that the $Mn_3$-$NH_3$ molecule is more stable. In order to prove this point, we calculated the coordination bond energy of these three kinds of $Mn_3$ molecules by density functional theory (DFT). The molecular structures of the three kinds of $Mn_3$ for our calculations are extracted from the corresponding experimental crystal structures. The single-point energies with different spin multiplicities are calculated by DFT at the B3LYP/6-31G** level, while atom Mn is treated with LANL2DZ pseudo potential. All the calculations are performed with Gaussian-09 program.

The results show that the Mn-O coordination bond energy of $Mn_3$-$CH_3OH$ and $Mn_3$-$H_2O$ are 0.54 eV and 0.57 eV, while the Mn-N coordination bond energy of $Mn_3$-$NH_3$ is 0.76 eV Thus, the large bond energy gap between Mn-N and Mn-O inhibits the backward reaction of $Mn_3$-$NH_3$ cannot take place, whereas the bond energy of $Mn_3$-$CH_3OH$ and $Mn_3$-$H_2O$ are quite close enabling mutual transformation. Fig.1c exhibits the mutual transformation process of these three kinds of $Mn_3$. As depicted in Fig.2, the molecular structure of $Mn_3$-$CH_3OH$, $Mn_3$-$H_2O$ and $Mn_3$-$NH_3$ are the same except for the ligands. Hence, the total spin of the molecule, which depends on the intramolecular structure, is expected to be the same for the three kinds of $Mn_3$. This is verified by DFT calculations which showed that the ground state spin for all these molecules is the same with $S$=6. Also, the spin densities exhibited a considerable localization on the $Mn^{3+}$ ions (see supplementary material). It thus indicates that the single molecular magnetism of the three kinds of $Mn_3$ are the same. On the other hand, the different ligand leads to a change of the lattice constants and the length of hydrogen bond (L). For $Mn_3$-$CH_3OH$, a=b=13.4446Å, c=34.4519Å, L=2.0225Å; for $Mn_3$-$H_2O$, a=b=13.1471Å, c=34.501Å, L=1.9657Å; for $Mn_3$-$NH_3$, a=b=12.9806Å, c=34.7392Å, L=2.0081Å. The small change of the length of hydrogen bond has a great influence on the macromagnetism and the QTM effect of the three kinds of $Mn_3$, as discussed below.

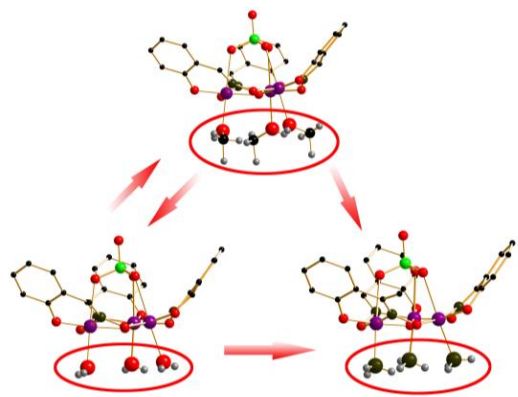

**Figure 2.** The mutual transformation process of these three kinds of $Mn_3$. The essence of the process is substitution of the ligands that are labeled by red circle. Colour code: Mn, purple; O, red; C, black; N, olive green; Cl, green; H, cyan.

To illustrate the effects of the different ligands on the magnetic properties, we performed direct current (dc), alternating current (ac) susceptibility and heat capacity on single-crystal of $Mn_3$-$CH_3OH$, $Mn_3$-$H_2O$ and $Mn_3$-$NH_3$, respectively. The dc magnetic measurements are performed on 7 T SQUID-VSM (Quantum Design) at 1.8 K. The single-crystal samples are first cut into the regular shape, then they are well oriented and fixed on a home-made Teflon cubic, which is glued on the sample holder. The magnetic easy axis is ensured to be parallel to the applied magnetic field. The ac magnetic measurements and heat capacity measurements are performed carried on 14 T PPMS (Quantum Design) equipped with standard ac magnetometer system option and standard heat capacity option.

Recent researches[25] show that $Mn_3$-$CH_3OH$ exhibit hysteresis loops with QTM effect and has no phase-transition due to the weak AFM intermolecular exchange interaction while the $Mn_3$-$H_2O$ exhibits an AFM phase transition at $T_N$=6.5 K. For $Mn_3$-$NH_3$, Fig. 3a exhibits the $M/H$-$T$ curves during field-cooling (FC) process. The magnetization first shows an ascent as temperature decreases then starts to drop at about 7.5 K, which suggests that there is an AFM phase-transition. To further determine the phase-transition temperature, we fall back on the heat capacity ($C_P$-$T$) curves in different magnetic fields shown in Fig. 3b. It is seen that the peak shifts to low temperature as the magnetic field is increasing, which is a feature of AFM phase-transition. And the phase-transition temperature is seen to be $T_N$=5.5 K at zero field which is consistent with the temperature where the $M/H$-$T$ curves drop steeply (see supplementary material). Fig. 3c shows the sketch map of the AFM structure of $Mn_3$. The AFM long range correlation is formed in ab plane. In c direction, there is no magnetic correlation between molecules. Thus, it can be regarded as a stack structure of independent AFM layers. According to the mean-field theory, the phase-transition temperature is determined by the strength of the intermolecular exchange interaction and the number of the nearest neighboring molecules. As mentioned above, the three kinds of $Mn_3$ have the same lattice structure, thus the phase-transition temperature should only proportional to the strength of intermolecular exchange interaction. Consequently, we can make a conclusion that the AFM intermolecular exchange interaction of $Mn_3$-$NH_3$ is stronger than $Mn_3$-$CH_3OH$ and weaker than $Mn_3$-$H_2O$.

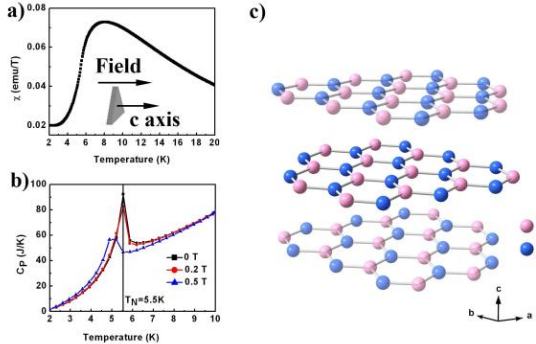

**Figure 3.** (a) The susceptibility vs temperature curves of $Mn_3$-$NH_3$ during field cooling process. The applied magnetic field is $H$=100 Oe and is parallel to c axis. (b) Heat capacity vs temperature curves of the same sample at different magnetic fields. (c) The sketch map of the AFM structure of $Mn_3$. The light pink and light blue balls represent molecules with spin down and spin up, respectively. The direction of the spin is parallel to the c axis.

To get more information of the AFM intermolecular exchange interaction of the three kinds of $Mn_3$, we performed a dc hysteresis experiment. Fig. 4a and Fig. 4b depict the normalized hysteresis loops and the derivative curves of these three kinds of $Mn_3$ at $T$=1.8K. It can be seen that all of the loops exhibit QTM steps. However, the shape of the loops and the position of the QTM peaks shown in Fig. 3b of them are different, indicating that the height of the anisotropy energy barrier and the strength of intermolecular AFM exchange interaction are different.

Additionally, recent work has reported that the $Mn_3$-$CH_3OH$ could transform into $Mn_3$-$H_2O$ when the sample is placed in air for about a month[27]. Hence, to check the stability of $Mn_3$-$NH_3$ in air, we performed the dc susceptibility experiments on the same single-crystal $Mn_3$-$NH_3$ sample after storing in air for a month. The hysteresis loops shown in Fig.4c suggest that the macromagnetism of $Mn_3$-$NH_3$ remains unchanged, which also means the lattice structure and the easy axis are unchanged. Besides, the position of the main QTM resonant fields shown in Fig.4d are the same indicating that the intermolecular AFM exchange interaction is invariant.

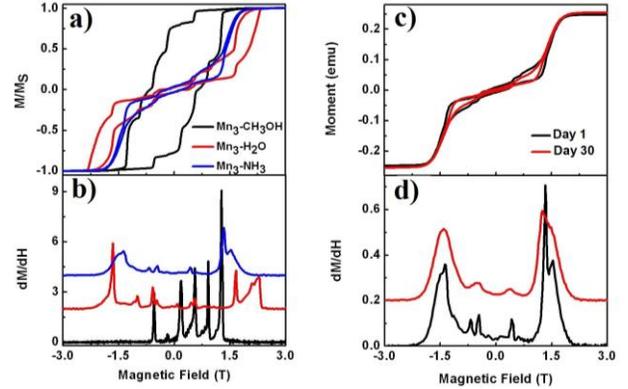

**Figure 4.** (a) Normalized magnetization hysteresis loops of $Mn_3$-$CH_3OH$ (black line), $Mn_3$-$H_2O$ (red line) and $Mn_3$-$NH_3$ (blue line). (b) The derivatives curves from -3 T to 3 T of the three kinds of $Mn_3$. They are shifted along y axis for clarity. c, The hysteresis loops of $Mn_3$-$NH_3$ at Day 1 and Day 30. d, The derivative curves from -3 T to 3 T of $Mn_3$-$NH_3$ at Day 1 and Day 30. The applied magnetic field is along c axis and the sweep field rate is 50 Oe/s.

Meanwhile, the ac susceptibility experiments of single-crystal of $Mn_3$-$NH_3$ were also performed. The in phase component curves shown in Fig. 5a first exhibit ascent as the temperature is decreased then drops at about 7.5 K, which is coincident with the results of dc magnetization experiments. Moreover, the peaks do not move with the change of frequency – a characteristic of AFM phase-transition. Meanwhile, it is seen that the peaks of out of phase components shown in Fig. 5b move to higher temperature as the frequency is increased, which is the typical characteristic for the spin-flip relaxation of SMMs. Using the Arrhenius law[28]: the effective energy barrier of $Mn_3$-$NH_3$ is figured out $U_{eff}$ = 46.85 K±1.91 K, with the fitting curve shown in the supplementary material. The effective energy barrier of $Mn_3$-$NH_3$ is larger than $Mn_3$-$H_2O$[26] and smaller than $Mn_3$-$CH_3OH$.

$$\tau = \tau_0 \exp(\frac{U_{eff}}{k_B T}) \qquad (1)$$

As just stated above, Mn$_3$-CH$_3$OH can be transformed into Mn$_3$-H$_2$O while Mn$_3$-H$_2$O can be changed into Mn$_3$-NH$_3$ when they are exposed in air or dried ammonia gas for a period of time. Hence, the coexistent state of Mn$_3$-CH$_3$OH and Mn$_3$-H$_2$O, Mn$_3$-H$_2$O and Mn$_3$-NH$_3$ should be achieved when we control the exposure time of Mn$_3$-CH$_3$OH and Mn$_3$-H$_2$O in air or dried ammonia gas respectively. Fig. 6a exhibits the out of phase component of Mn$_3$-CH$_3$OH sample exposed in air for 6 days. It is clear that there are two distinguishable peaks in the curves (see supplementary material) and both of them move to higher temperature as frequency increasing. It indicates that there are two spin-flip relaxation processes in the system which is a proof that Mn$_3$-CH$_3$OH and Mn$_3$-H$_2$O molecules coexist in the same sample. Meanwhile, Fig. 6b shows the out of phase component of Mn$_3$-H$_2$O sample which is exposed in dried ammonia gas for 6 days. An enveloping line can be observed. Because the $U_{eff}$ of Mn$_3$-H$_2$O[26] and Mn$_3$-NH$_3$ is close to each other, the enveloping line should be formed by two unimodal curves (see supplementary material) overlapped which belong to the spin-flip relaxation process of Mn$_3$-H$_2$O and Mn$_3$-NH$_3$, respectively. Therefore, the Mn$_3$-CH$_3$OH and Mn$_3$-H$_2$O, Mn$_3$-H$_2$O and Mn$_3$-NH$_3$ can be coexistent in the same sample if we appropriately control the time that the sample is exposed in air or in ammonia gas.

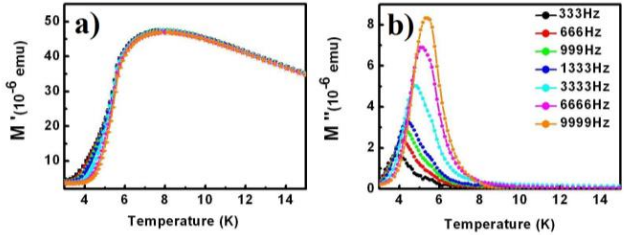

**Figure 5.** (a), The in phase component vs temperature curves of Mn$_3$-NH$_3$ at different frequencies. (b), The out of phase component vs temperature curves of Mn$_3$-NH$_3$ at different frequencies.

In order to quantitatively analyze the magnetic properties of these three kinds of Mn$_3$, we utilize the equation of QTM fields of SMM with identical intermolecular exchange interaction from $|-S\rangle$ to $|S-l\rangle$ described as[27]:

$$H_Z = lD/g\mu_0\mu_B + (n_\downarrow - n_\uparrow)JS/g\mu_0\mu_B \qquad (2)$$

where $D$ is the magnetic anisotropy parameter[21,22,25,27] leading to the energy barrier $U=DS^2$, $J$ is antiferromagnetic intermolecular exchange interaction parameter, $n\downarrow$ and $n\uparrow$ stand for the number of a tunneling molecule's neighboring molecules occupying the $|-S\rangle$ and $|S\rangle$ state, respectively. $S$ represents the ground state spin, $g$ is Lande factor, $\mu_B$ is Bohr magneton, $\mu_0$ is permeability of vacuum. Using this equation and the specific position of the QTM peaks shown in Fig. 2b we can figure out the value of $D$ and $J$ of SMMs[27]. For Mn$_3$-CH$_3$OH, $D$=0.98 K, $J$=-0.041 K[27]; for Mn$_3$-H$_2$O, $D$=0.925 K, $J$=-0.132 K[25]; and for Mn$_3$-NH$_3$, $D$=0.884 K, $J$=-0.094 K. The results indicate that the anisotropy energy barrier ($U=DS^2$) of Mn$_3$-NH$_3$ is the smallest which is in contradiction with the effective energy barrier $U_{eff}$ given by ac magnetic experiment. And it should also be noted that the value of the $U_{eff}$ is bigger than $U$. The reason is that $U=DS^2$ is the single particle energy barrier, but, there is the multi-body spin-flip process[29] in these systems due to the existence of intermolecular exchange interaction. The multi-body spin-flip process will significantly increase the height of the effective energy barrier; however, for Mn$_3$-H$_2$O and Mn$_3$-NH$_3$ the intermolecular exchange interaction is strong enough to lead to the AFM phase transition which will suppress the multi-body spin-flip process. Thus, the height of the effective energy barrier will decrease as the intermolecular exchange interaction increasing.

On the other hand, the AFM intermolecular exchange interaction of Mn$_3$-H$_2$O and Mn$_3$-NH$_3$ is about three times and two times stronger than Mn$_3$-CH$_3$OH, respectively. Hence, the AFM phase-transition only can be observed in these two systems. As mentioned above, the intermolecular exchange interaction depends on the hydrogen bond between molecules. The results show that the relationship between the strength of the interaction and the length of hydrogen bond is monotonic as shown in Fig. 3c, the shorter the length of hydrogen bond the stronger is the interaction. The reason is that the intermolecular exchange interaction is formed by super-exchange pathways[30] which is determined by the overlap integral of the wave function. Shorter distance leads to larger overlapping of the wave function resulting in bigger exchange interaction parameter $J$[31,32].

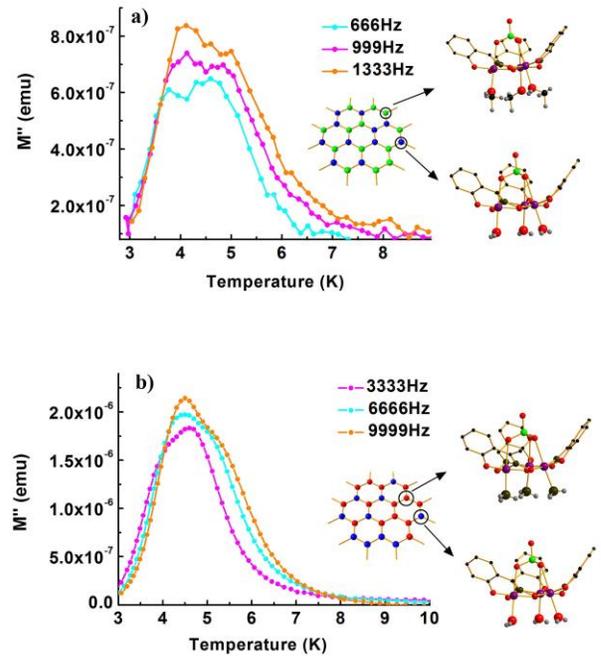

**Figure 6.** (a), The out of phase component vs temperature curves of Mn$_3$-CH$_3$OH and Mn$_3$-H$_2$O coexistent state at different frequencies. (b), The out of phase component vs temperature curves of Mn$_3$-NH$_3$ and Mn$_3$-H$_2$O coexistent state at different frequencies.

In summary, we have achieved the mutual transformation of ligand in monocrystal single-molecule magnet $Mn_3$ by external stimuli (different chemical gas atmosphere). The methanol ligands on $Mn_3$-$CH_3OH$ can be substituted by water ligands when the sample is exposed in air for a few days. Interestingly, the water ligands on $Mn_3$-$H_2O$ can also be replaced by ammonia using the same method. Importantly, the hydrogen bond between molecules depends on these ligands which determine the intermolecular exchange interaction. As a result, these three kinds of $Mn_3$ exhibit different dc hysteresis loops, QTM effect and ac spin-flip effective energy barrier at low temperature. Moreover, $Mn_3$-$H_2O$ and $Mn_3$-$NH_3$ exhibit an AFM phase transition at $T_N$= 6.5K and $T_N$= 5.5K respectively, that is relative high in the field of transition metal SMMs. On the other hand, we also obtain the coexistent state of $Mn_3$-$CH_3OH$ and $Mn_3$-$H_2O$, $Mn_3$-$H_2O$ and $Mn_3$-$NH_3$ when we control the time that the sample exposed in air or in ammonia gas, that indicates the magnetic properties of $Mn_3$ is tunable. More importantly, the single-crystal structure of $Mn_3$ keeps unchanged during the whole transformation process. Therefore, our results open up an avenue for exploring the non-destructive production of two-dimensional SMM-network by external stimuli. We also believe that the gas-responsive $Mn_3$ will have great potential in the multifunctional magnetic material.

See supplementary materials for the results of dc and ac magnetic measurements of three kinds of $Mn_3$ and corresponding theoretical analysis.

Author to whom correspondence should be addressed. E-mails : luojj@ime.ac.cn

This work was financially supported by the National Natural Science Foundation of China (No. 61404161). The authors gratefully acknowledged S. S. K. for the assistance with magnetic measurements at Institute of Physics Chinese Academy of Sciences.

## REFERENCES

(1) L. Bogani, W. Wernsdorfer, *Nat. Mater.* **7**, 179(2008).
(2) A. R. Rocha, V. M. García-suárez, S. W. Bailey, C. J. Lambert, J. Ferrer, S. Sanvito, *Nat. Mater.* **4**, 335(2005).
(3) L. Zhu, K. L. Yao, Z. L. Liu, L. Lu, T. Zheng, Q. Wu, A. M. Schneider, D. Zhao, L. Yu, *Appl. Phys. Lett.* **96**, 082115(2010).
(4) C. Romeike, M. R. Wegewijs, H. Schoeller, *Phys. Rev. Lett.* **96**, 196805(2006).
(5) M. Matteo, P. Francesco, S. Philippe, D. Chiara, O. Edwige, S. Corrado, M. T. Anna, A. Marie-Anne, C. Andrea, G. Dante, S. Roberta, *Nat. Mater.* **8**, 194(2009).
(6) C. W. Zhang, H. Kao, J. M. Dong, *Phys.Lett. A* **373**, 2592(2009).
(7) T. Dietl, *Nat. Mater.* **9**, 965(2010).
(8) R. R. Nair, M. Sepioni, I-L. Tsai, O. Lehtinen, J. Keinonen, A. V. Krasheninnikov, T. Thomson, A. K.Geim, I. V. Grigorieva, *Nat. Phys.* **8**, 199(2012).
(9) O. V. Yazyev, L. Helm, *Phys.* Rev. B **75**, 125408(2007).
(10) J.Cervenka, M. L. Katsnelson, C. F. Flipse, *Nat. Phys.* **5**, 840(2009).
(11) T. C. Stamatatos, K. A. Abbound, W. Wernsdorfer, G. Christou, *Angew. Chem. Int. Ed.* **46**, 884(2007).
(12) Y. Suzuki, K. Takeda, K. Awaga, *Phys. Rev. B* **67**, 132402(2003).
(13) R. Bircher, G. Chanoussant, C. Dobe, H. U. Güdel, S. T. Ochsenbein, A. Sieber, O. Waldman, *Adv. Funct. Mater.* **16**, 209(2006).
(14) A. Prescimone, C. J. Milios, J. Sanchez-Benitez, K. V. Kamenev, C. Loose, J. Kortus, M. Moggach, M. Murrie, J. E. Warren, A. R. Lennie, S. Parsons, E. K. Brechin, *Dalton Trans.* **25**, 4858(2006).
(15) Y. Cui, Y. Wu, Y. R. Li, R.Y. Liu, X. L. Dong, Y. P. Wang, *S Sci. China-Phys. Mech. Astron.* **57**, 1299(2014).
(16) F. B. Meng, Y-M. Hervault, Q. Shao, B. H. Hu, L. Norel, S. Riguat, X. D. Chen, *Nat. Commun.* **5**, 3023(2013).
(17) H. Miyasaka, M. Yamashita, *Dalton Trans.* 399 (2007).
(18) R. Inglis, G. S.Papaefstathiou, W. Wernsdorfer, E. K. Brechin, *Aust. J. Chem.* **62**, 1108(2009).
(19) R. Inglis, S. M. Taylor, L. F. Jones, G. S. Papaefstathiou, S. P. Perlepes, S. Datta, S. Hill, W. Wernsdorfer, E. K. Brechin, *Dalton Trans.* 9157(2009).
(20) R. Inglis, L. F. Jones, G. Karotsis, A. Collins, S. Parsons, S. P. Perlepes, *Chem. Commun.* 5924(2008).
(21) L. Thomas, F. M. Lionti, R. Ballou, D. Gatteschi, R. Sessoli, B. Barbara, *Nature* **383**, 145(1996).
(22) J. R. Friedman, M. P. Sarachik, J. Tejada, R. Ziolo, *Phys.Rev. Lett.* **76**, 3830(1996).
(23) G. Q. Bian, T. Kuroda-Sowa, H. Konaka, M. Hatano, M. Maekawa, M. Munakata, H. Miyasaka, M. Yamashita, *Inorg. Chem.* **43**, 4790 (2004).
(24) W. Wernsdorfer, N. A. Alcalde, D. N. Hendrickson, G. Christou, *Nature* **416**, 406 (2002).
(25) Y. Cui, Y. R. Li, R. Y. Liu, Y. P. Wang, *arXiv:1501.05484* (2015).
(26) Y. Cui, Y. R. Li, R. Y. Liu, Y. P. Wang, *Chin. Phys. B* **23**, 067504(2014).
(27) Y. R. Li, R. Y. Liu, H. Q. Liu, Y. P. Wang, *Phys. Rev. B*, **89**, 184401(2014).
(28) F. Lius, J. Bartolomé, J. F. Fernández, J. Tejada, J. M. Hernández, X. X. Zhang, R. Ziolo, *Phys. Rev. B* **55**, 11448(1997).
(29) W. Wernsdorfer, S. Bhaduri, R. Tiron, D. N. Hendrickson, G. Christou, *Phys. Rev. B* **72**, 214429(2005).
(30) H. A. Kramers, *Physica* **1**, 182(1934).
(31) P. W. Anderson, *Phys. Rev.* **79**, 350(1950).
(32) J. B. Goodenough, *Phys. Rev.* **100**, 564(1955).